\documentclass[conference]{IEEEtran}
\usepackage{blindtext, graphicx}
\usepackage{draftwatermark}
\SetWatermarkText{PREPRINT}
\SetWatermarkScale{1}
\ifCLASSINFOpdf
\else
\fi
\begin{document}
%
\title{Analysis of Blink Rate Variability during reading and memory testing}


\author{\IEEEauthorblockN{
Temesgen Gebrehiwot,
Rafal Paprocki and
Artem Lenskiy}

\IEEEauthorblockA{Korea University of Technology and Education\\
1600, Chungjeol-ro, Byeongcheon-myeon,
Dongnam-gu, Cheonan-si, Chungcheongnam-do 31253,\\ Republic of Korea\\
email: rafal.paprocki@gmail.com, lensky@koreatech.ac.kr}
}


%


\maketitle

\begin{abstract}
In this paper we investigated how statistical properties of the blink rate variability changes during two mental tasks: reading a passage and memory testing. To construct time series of inter-blink intervals (blink rate variability) we detected exact blink time in EEG recordings using our blink detection algorithm. We found that among 13 subjects, all subjects blinked less during reading session. Moreover, standard deviation of the blink rate variability is higher during reading. Thus, we conclude that the variability of inter-blink intervals decreases during tasks that require concentration and intense mental activity. 
\end{abstract}


%
\IEEEpeerreviewmaketitle

\section{Introduction}
Blinking is a semi-autonomic closing of the eyelids. Occurrence, reasons and characteristics of blink rate, varies between animals, and it is possible to trace the evolution of the blinking mechanisms through species. \cite{1}. Every blink is associated with a spread of oils and mucous secretions across the surface of the eye to keep eyes from getting dry and disinfect them. Reduced blink rate causes eye redness and dryness also known as Dry Eye, which belongs to the major symptoms of the Computer Vision Syndrome \cite{2}. Besides blink keeps eyes protected against potentially damaging stimuli, such as bright lights and foreign bodies like dust.\\
However, people don't even notice the world going into darkness every few seconds. The sudden changes in an image due to saccades or blinks, do not interfere with our subjective experience of continuity, the very act of blinking suppresses an activity in several areas of the brain responsible for detecting environmental changes, so that one experience the world as continuous. \\
The blinks have been known to be linked to interior brain activities. Ponder and Kennedy implicated that high processes are major determinants of Blink enhancements and inhibition \cite{4}. Moreover, it has been reported the synchronous behavior in blinking between listener and speaker in face-to-face conversation \cite{3}.\\
Researchers have shown that blinks can play a significant role in detecting many different brain disorder and brain activities. Spontaneous Blink Rate (BR) has been studied in many neurological diseases like Parkinson's disease and Tourette syndrome \cite{5}, \cite{6}, \cite{7}. The use of blink detection does not stop there. Researchers have found that blink rates can be used as a source of data in detecting psychiatric disorders like schizophrenia and attention hyperactivity all this is because blinks are regarded as non-invasive peripheral markers of the central dopamine activity which makes their accurate detection more important.\\
It is evident that eye blinks in general and the BR particularly is linked to brain activity. In this paper, we investigate the relationship between the blink rate variability and two activities reading a passage and testing memory. To estimate blink rate variability, we utilize a blink detection algorithm that we proposed earlier \cite{8} and briefly describe in section 3. The experimental setup is described in section 2. In the sections 4 we discuss on the relationship between the blink rate and the task, and in section 5 we elaborate on the statistical properties of blink rate variability and their relationship to the reading task and memory test.    

\section{Experimental setup}
\subsection{Data acquisition}
The video stream while subjects were performing the test was captured with a Pointgrey Flea3 high frame rate USB camera. We stored the video for future work. Simultaneously, EEG signals were recorded using Mitsar-EEG 201 amplifier and accompanying WinEEG software. The electrodes were placed according to the international “10-20 system” standards of electrode placement \cite{9}. Electro-gel was injected into electrodes hollow in order to decrease the electrode-skin resistance. Currently, we are more focused on the recording of the blinks than the brain data, but In the future we are planning to analyze EEG to detect different types of brain activities while performing these exams.

\begin{figure}[!htbp]
\includegraphics[width=3.5in]{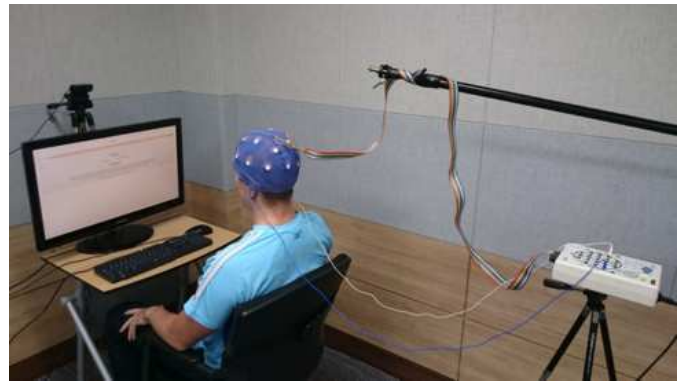}
\caption{Experimental setup}
\label{fig1}
\end{figure}

\subsection{Testing procedure}
Eighteen young subjects (15 men and 3 women) aged from 19-25 years, were recruited for the experiment. All provided their written consent. The subjects had no history of psychiatric illness and they had not been affected by any significant medical, neurological or ophthalmological illness. To avoid substance abuse, we prepared a questionnaire testifying that they had none of these issues. Among them, five subjects were dropped due to heavy noise caused by the subjects falling asleep, adjusting the cap or constant head movements. \\
The test for the data recording session consisted of two different stages: (a) reading passage, and (b) comprehension test. The testing software was developed in Java. It was designed in such a way that it does not require any interventions from either the subjects or the experiment supervisor to conduct this experiment. The whole testing session took 10 minutes. The first 5 minutes is the passage reading session. The passage consists of basic facts about Ethiopia, and the target of this passage was to record data while subjects were trying to comprehend the passage. Lastly, subjects were given a 5 minutes comprehension test about the passage. The questions were derived from the information provided in the passage. It was designed to get subject to retrieve the facts from the memory. Since EEG is prone to noise due to movement, we made sure that subjects answered the question by using the numbers from 1 to 6 on the keypad.  

\begin{figure}[!htbp]
\includegraphics[width=3.5in]{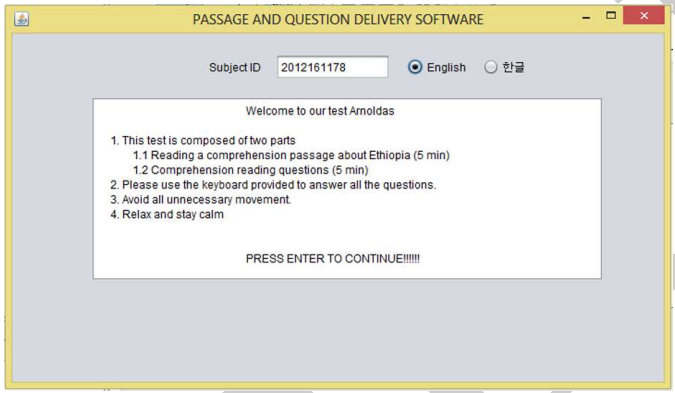}
\caption{Question delivery software}
\label{fig2}
\end{figure}

\section{Estimation of the Blink Rate Variability}
EEG signals were recorded while participants were taking the tests. We used bipolar montage while recording these signals, i.e. signals are the potentials between Fp1 and Fp3 electrodes, and Fp2 and Fp4 electrodes. The recordings were then imported in the form of CSV files to Matlab for further analysis. The process of blink detection is described in our previous work \cite{8}. Here we briefly outline it.\\
Generally, the blink detection is divided into two parts: the preprocessing part and the blink detection part. The preprocessing part consists of the following steps: (a) bandpass filtering, (b) cut extreme amplitudes using estimated cumulative distribution function, characterizing signal amplitudes distribution, (c) independent component analysis and (d) selection of the independent component with eye blinks. The blink detection part of the detection algorithm consists of (d) signal thresholding, (e) candidate extraction, (f) polynomial fitting with finding maximum in the polynomial function (Fig. \ref{fig4}), and (g) finally calculating blink rate variability (Fig. \ref{fig5}).

\begin{figure}[!htbp]
\includegraphics[width=3.5in]{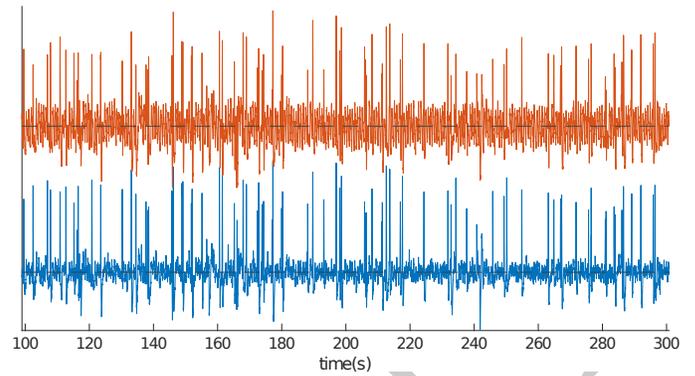}
\caption{Fp1-Fp3 and Fp2-Fp4 electrode pairs}
\label{fig3}
\end{figure}

\begin{figure}[!htbp]
\includegraphics[width=3.5in]{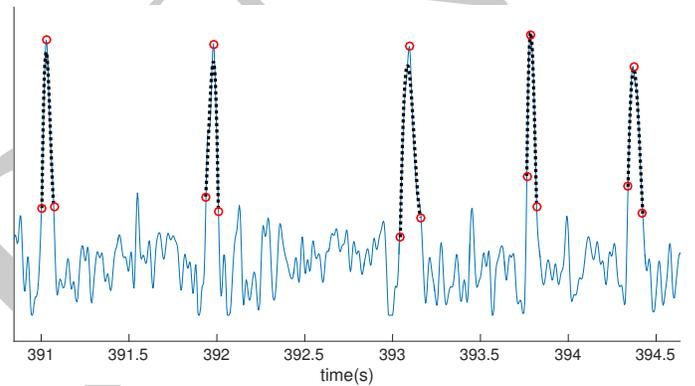}
\caption{3rd degree polynomial function is fitted to every waveform blink candidate. The localized peak corresponds to the time of a blink.}
\label{fig4}
\end{figure}

\begin{figure}[!htbp]
\includegraphics[width=3.5in]{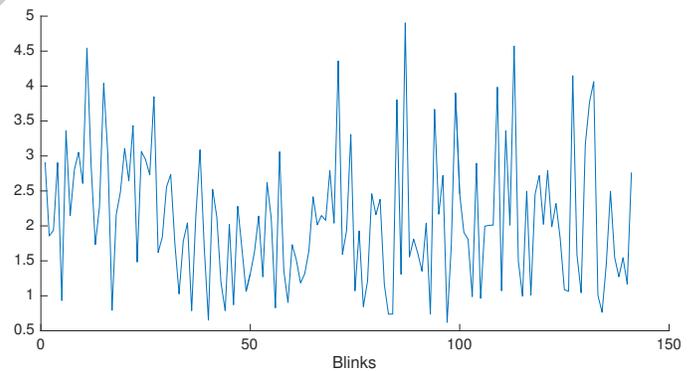}
\caption{Extracted blink rate variability}
\label{fig5}
\end{figure}

\section{Blink rate and blink rate variability during reading and memory testing}
\begin{figure}[!htbp]
\includegraphics[width=3.5in]{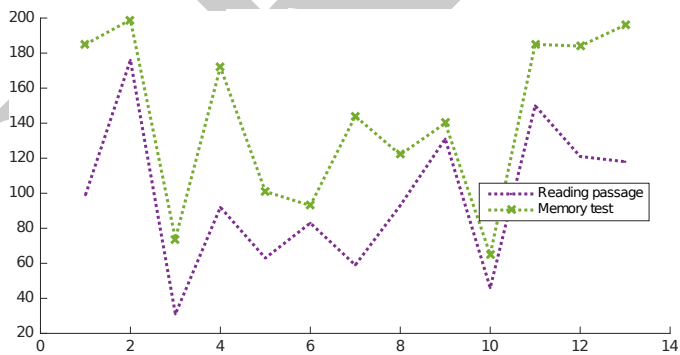}
\caption{Number of blinks depending on the task}
\label{fig6}
\end{figure}

In the work by Bentivoglio et al, BR during rest, reading and conversation was studied. The authors investigated the blink rate (BR) patterns; they showed that the patterns influenced more by cognitive processes rather than by age, eye color or gender \cite{10}. Hart H.W distinguishes three types of an eye blink: spontaneous, reflex and voluntary\cite{11}. He also estimated that blinks occur on average, about 12 -15 times/min \cite{11}. It was noted that spontaneous BR increases at the evening time \cite{12}. The study of BR at the evening time is especially important due to the fact that the ninth cause of death globally are car accidents according to World Health Organization (WHO) and the big part of the accidents are caused by the drowsiness of drivers. According to National Motor Vehicle Crash Causation Survey (NMVCCS) 30\% of car accidents occur in the evenings \cite{13}. It is known that workload increases heart rate and heart rate are known to decrease in monotonous and drowsy conditions \cite{14}. BR is inversely correlated with increase of workload, so blinks can be used to detect drowsiness before bad consequences occur \cite{14}.\\
We investigate how number of blinks changes depending on the task. We considered two tasks: reading and memory test based on the text that subjects were given. Among 13 subjects, all blinked less while reading. Holland et. al reported that blinking is related to certain cognitive processes. The authors state that “blinking is a function of memory load, such that the more items in memory, the fewer the blinks”.  Reading is a process in which the eyes quickly move to assimilate text. It is necessary to understand visual perception and eye movement in order to understand the reading process. Reading involves a prolonged focus on reading material as each words is related to its predecessors to make sense of the whole sentence.\\
Operational memory, which is used during solving mental tasks, and visual imagination, may share components with a visual perceptual system. To avoid interference of cognitive processes, blinking is slowed down \cite{16}. Number of blinks per each of 13 subjects is presented in Fig.\ref{fig6}. 

\begin{table}[]
\centering
\caption{Number of blinks of subjects oer sessions, calculated by our algorithm}
\label{my-label}
\begin{tabular}{l|ll}
\hline

\begin{tabular}[c]{@{}l@{}}Subject \end{tabular} & 
\begin{tabular}[c]{@{}l@{}}Passage\end{tabular} & 
\begin{tabular}[c]{@{}l@{}}Test\end{tabular}  

\\ \hline
1 &	    100 &	    186 \\
2 &	    177 &	    200 \\
3 &	    32 &	    75 \\
4 &	    93 &	    173 \\
5 &	    64 &	    102 \\
6 &	    84 &	    94 \\
7 &	    60 &	    145 \\
8 &	    94 &	    123 \\
9 &	    132 &	    141 \\
10 &	    47 &	    66 \\
11 &	    151 &	    186 \\
12 &	    122 &	    185 \\
13 &	    119 &	    196 \\
\hline
\end{tabular}
\end{table}

\begin{figure}[!htbp]
\includegraphics[width=3.5in]{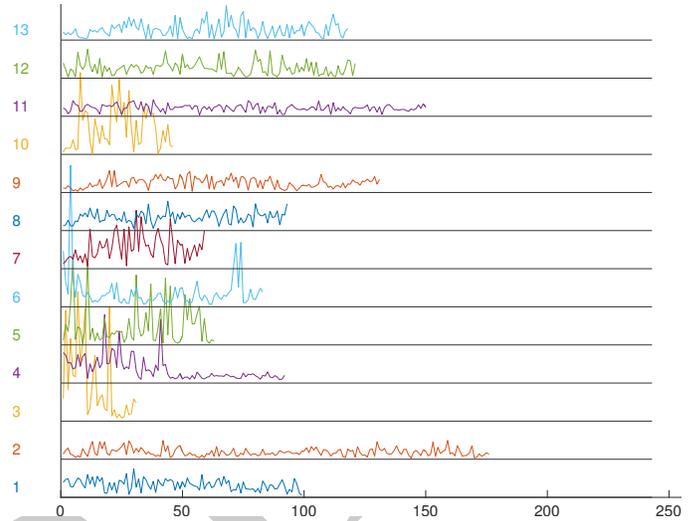}
\caption{Blink rate variability during passage reading}
\label{fig7}
\end{figure}

\begin{figure}[!htbp]
\includegraphics[width=3.5in]{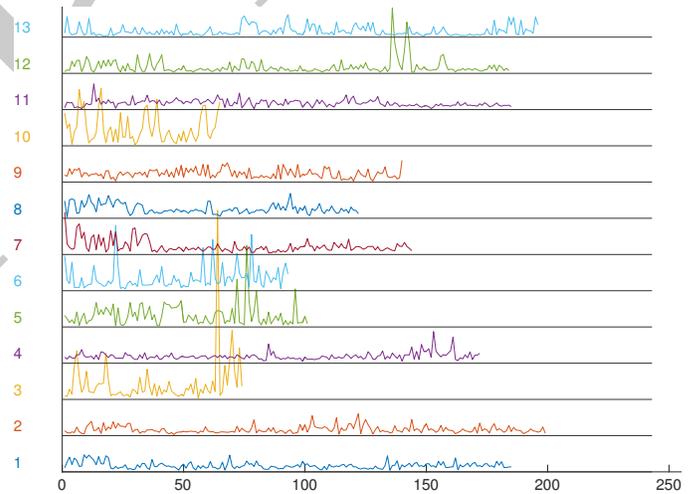}
\caption{Blink rate variability during memory testing}
\label{fig8}
\end{figure}

Fig. \ref{fig7} and Fig. \ref{fig8} show BRV for all of the subjects during reading and testing comprehension. The X-axis is the blink interval where Y-axis represents interval lengths per each subject. Even though BRV visually looks different, the statistical properties are consistent. The mean and the standard deviations of BRV are lower during memory test than during reading session for all subjects.

\begin{figure}[!htbp]
\includegraphics[width=3.5in]{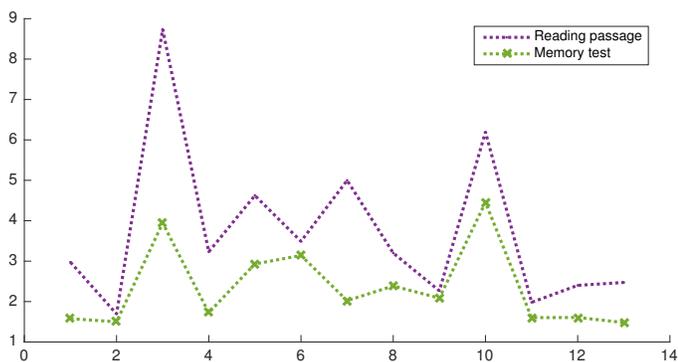}
\caption{Average inter-blink.}
\label{fig9}
\end{figure}

\begin{figure}[!htbp]
\includegraphics[width=3.5in]{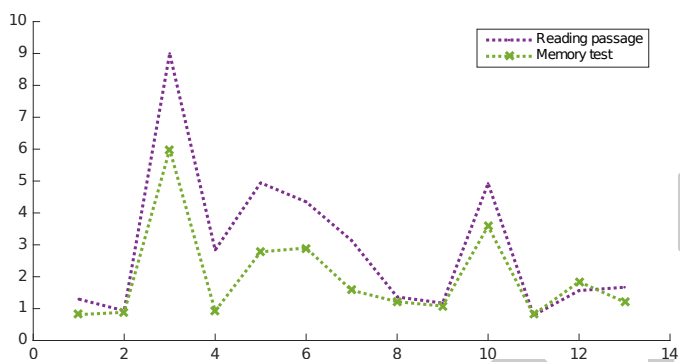}
\caption{Standard deviation of IBI dynamics}
\label{fig10}
\end{figure}

Comprehension test, as the task involving memory, increases BR, while reading reduces it. Reading text that contains new information appears to be a heavier mental process than memory testing, and operational memory is more in use. Moreover, logical constructions in the text are longer, and ideas are more developed than in the test. Therefore, there could be a reflection in concept of blinking as an interlude between ideas or sentences \cite{16}. \\
Time between blinks changes, for reading it is shorter and elongates for test. Fig. \ref{fig9} presents average inter-blink per each subject in both stages, reading and memory test. Fig. \ref{fig10} shows standard deviation of IBI dynamics.  Longer inter-blink in reading is caused because of fixation points while reading. Humans read each word at a time this is called fixation. From one fixation to another there is a search for the next fixation so this requires less blinks and more focus.  Also longer inter-blinks are caused by the fact that reading is a cognitive process that increases brain activity \cite{17}.

\section{Conclusion}

In 1927 Ponder and Kennedy reported that high processes are major determinants of blink enhancements and inhibition. They also mentioned that blinks might serve as an index of attention, or as they termed it “mental tension”. They inferred that person inhibits blinking while actively engaged in information abstraction. Interesting results were presented by Hall, who concluded that blinks do not occur randomly during reading.  \\
Our results have shown that there is lesser number of blinks in reading a passage than performing a memory test. Reading has lower blink rates because people read sentences by fixating on words, which results in a decrease of blinking rates. Our findings supports\cite{17} that blink behavior during reading is under perceptual and cognitive control. Even though it was a different experiment set up our finding is concurrent with the findings of \cite{18} that reports on a significant and a positive correlation between inter blink intervals and subsequent memory encoding. Our results show higher blink rates while subjects performed the memory test. In summary our experiment shows that inter blink intervals have strong correlation with the task a person is performing. Our findings can lead to further studies on the correlation between IBI’s and different human activities.

\ifCLASSOPTIONcaptionsoff
  \newpage
\fi



%

%

\begin{IEEEbiography}[{\includegraphics[width=1in,height=1.25in,clip,keepaspectratio]{picture}}]{John Doe}
\blindtext
\end{IEEEbiography}




\end{document}